\documentclass[journal]{IEEEtran}
\usepackage{cite}

\ifCLASSINFOpdf
  \usepackage[pdftex]{graphicx}
\else
\fi
\usepackage{amsmath}
\usepackage{amsfonts}
\usepackage{amssymb}

\usepackage{array}

\usepackage[dvipsnames]{xcolor}
\usepackage{tikz}
\usetikzlibrary{arrows.meta}
\usetikzlibrary{patterns}
\usepackage{bbm}
\tikzstyle{line} = [draw, -{Latex[length=3mm,width=1.5mm]}]
\makeatletter
\def\@makefnmark{%
  \leavevmode
  \raise.9ex\hbox{\fontsize\sf@size\z@\normalfont\tiny\@thefnmark}}
\makeatother

\usepackage{glossaries}

\newacronym{3gpp}{3GPP}{3rd Generation Partnership Project}
\newacronym{bs}{BS}{base station}
\newacronym{cdf}{CDF}{cumulative distribution function}
\newacronym{ue}{UE}{user equipment}
\newacronym{los}{LoS}{line-of-sight}
\newacronym{mdt}{MDT}{minimization of drive tests}
\newacronym{ofdm}{OFDM}{orthogonal frequency division multiplexing}
\newacronym{tdoa}{TDOA}{time difference of arrival}
\newacronym{toa}{TOA}{time of arrival}
\newacronym{urllc}{URLLC}{ultra-reliable low-latency communications}
\newacronym{mar}{MAR}{maximum achievable rate}
\newacronym{pcr}{PCR}{probably correct reliability}

\newcommand{\mbf}[1]{\mathbf{#1}}
\newcommand{\mbs}[1]{\boldsymbol{#1}}
\newcommand{\mR}{\mathbb{R}}
\newcommand*{\cond}{\hspace*{1pt} |\hspace*{1pt}}
\newcommand{\norm}[1]{\left\lVert#1\right\rVert}
\newcommand{\mC}{\mathbb{C}}
\newcommand{\T}{\textsf{T}}
\renewcommand{\H}{\textsf{H}}

\DeclareMathOperator{\Var}{Var}

\newcommand{\e}{\text{e}}

\newcommand{\rev}[1]{#1}

\begin{document}
%
\title{A Primer on the Statistical Relation between Wireless Ultra-Reliability and Location Estimation}
%
%
%

\author{Tobias~Kallehauge,
        Pablo~Ramírez-Espinosa,
        Kimmo~Kansanen,~Henk~Wymeersch~\IEEEmembership{Senior Member,~IEEE}, and Petar~Popovski,~\IEEEmembership{Fellow,~IEEE}
\thanks{This work was partially supported by the Villum Investigator Grant in Denmark.}
\thanks{T. Kallehauge, P. Ramírez-Espinosa and P. Popovski are with the Department of Electronic Systems, Aalborg University, Denmark (e-mail: tkal@es.aau.dk; pres@es.aau.dk; petarp@es.aau.dk).}
\thanks{K. Kansanen is with the Norwegian University of Science and Technology, Trondheim, Norway (e-mail: kimmo.kansanen@ntnu.no).}%
\thanks{H. Wymeersch is with Chalmers University of Technology, Gothenburg, Sweden (e-mail: henkw@chalmers.se).}
}

%
%


\maketitle

\begin{abstract}
Location information is often used as a proxy to infer the performance of a wireless communication link. Using a very simple model, this letter unveils a basic statistical relation between the location estimation uncertainty and wireless link reliability. First, a Cram\'er-Rao bound for the localization error is derived. Then, wireless link reliability is characterized by how likely the outage probability is to be above a target threshold. We show  that the reliability is sensitive to location errors, especially when the channel statistics are also sensitive to the location. Finally, we highlight the difficulty of choosing a rate that meets target reliability while accounting for the location uncertainty.
\end{abstract}



%

\vspace{-0.2cm}

\section{Introduction} \label{sec:intro}
\IEEEPARstart{U}{}ser localization and \gls{urllc} are ubiquitous concepts in 5G networks \cite{Ghosh2019}. Reliability of wireless transmission is related, among others, to the behavior of the propagation channel, which is inherently correlated with spatial location. Consequently, exploiting this relation is envisioned as a promising direction in mobile networks, e.g., using location information to assist millimeter-wave communications  \cite{Maschietti2017}, the generation of channel maps for increased reliability and predictive resource allocation  \cite{Anese2011, risk_URLLC}, and channel charting for user localization \cite{Studer2018}. The standardization by \gls{3gpp} of \gls{mdt}\cite{3gpp_MDT} is an additional motivation, allowing the operators to utilize end-user devices measurements for the previously mentioned tasks.

In contrast to more conventional approaches where samples are acquired over time to estimate channel statistics and thus reliability \cite{marko}, the relation between channel and location brings forward the idea of using location to infer channel statistics and, ultimately, as a proxy for guaranteeing reliability in \gls{urllc}. 
Considering the latency introduced by estimating channel statistics, a communication system that predicts reliability based on localization using only a few measurements, is an attractive alternative. However, these reliability-guaranteeing methods would rely, among other aspects, on the ability to estimate location accurately, which raises the question: \emph{How can the accuracy of the localization procedures impact the wireless reliability guarantees?}

This letter will analyze a simplified framework to reveal the \emph{fundamental} relations between location uncertainty and reliability guarantees, neglecting other sources of uncertainty such as channel estimation. 
A \gls{ue} is intended to communicate with a \gls{bs}, and to isolate the impact of location uncertainty on reliability, the \gls{ue} is assumed to perfectly know the statistics for channel propagation at all locations around the \gls{bs}. If the location were perfectly known, the UE would correctly allocate resources to guarantee some level of reliability. However, given the uncertainty of the estimated location, the predicted reliability is also uncertain, and the \gls{ue} must account for scenarios where, e.g., the signal level is weaker at the true location than at the estimated location.

In this letter, we investigate \rev{the impact of location uncertainty in reliability} by modeling a system in the physical communication layer with the additional simplification of only considering one dimension for localization. This allows us to extract important conclusions without dealing with the complexity inherent to higher dimensional and higher layer cases. The localization performance is characterized through Fisher information analysis \cite{equivalent_fisher}, and reliability is statistically characterized following the \gls{pcr} approach in \cite{marko}.

\emph{Notation:} $\mathfrak{R}(z)$ and $\mathfrak{I}(z)$ are the real and imaginary parts $z$, and $\jmath$ is the imaginary unit. $(\cdot)^\T$ and $(\cdot)^\H$ are the matrix transpose and conjugate transpose, and $\norm{\cdot}$ is the $\ell_2$-norm. For matrix $\mbf{A}$, the submatrix with row $i$ to $j$ and column $k$ to $p$ is denoted $\mbf{A}_{i:j,k:p}$. $\mathcal{N}(\mu,\sigma^2)$ and $\mathcal{CN}(\mu,\sigma^2)$ denote Gaussian and complex circular symmetric Gaussian distributions with mean $\mu$ and variance $\sigma^2$. Finally, $E[\cdot]$ and $\Var[\cdot]$ denote, respectively, the expectation and the variance operators. 

\section{System model}
\subsection{Communication system and channel model}

We consider a simple 1-D framework with two \glspl{bs} at locations $x_1, x_2 \in \mR$ which communicate with a \gls{ue}  at location $x \in \mR$. Both \glspl{bs} and the user are equipped with a single antenna. An \gls{ofdm} modulation scheme is considered, with bandwidth $W$ and $N$ subcarriers spaced $\Delta_f = W/N$.  


The channel between the \gls{ue} and the \gls{bs} $i \in \{1,2\}$ is assumed to follow a two-path model with complex channel coefficient $a_{i,k}$ and associated delay $\tau_{i,k}$ for path $k$. The first path ($k=1$) characterizes the \gls{los} link, being thus deterministic and geometrically-dependent as \cite{tse_vis,wymeersch_beyond_2020}
\begin{align}
a_{i,1} = \sqrt{\frac{\lambda^2}{16\pi^2d_i^2}}\e^{-\jmath 2\pi d_i / \lambda} = \sqrt{P_L(d_i)} \e^{-\jmath\phi(d_i)},    
\end{align}
where $\lambda$ is the wavelength and $d_i = \norm{x - x_i}$. Naturally, it follows that $\tau_{i,1} = \norm{x-x_i}/c $ with $c$ the speed of light.
The second path ($k=2$) represents the contribution of the scattered paths, which cannot be mutually resolved and hence $a_{i,2}\sim\mathcal{CN}(0, \sigma^2_i(\Delta\tau_{i}))$  
with variance according to an exponential power delay profile \cite{krouk}:
\begin{align} 
\Var[a_{i,2}] = \sigma_i^2(\Delta\tau_i) =\frac{P_{L}(d_i)}{\rho}\exp\left(- \frac{\Delta\tau_i}{\rho} \right), \label{eq:scatter_stats}
\end{align}
where $\Delta\tau_i = \tau_{i,2} - \tau_{i,1}$ is the excess delay and $\rho > 0$ controls how fast the power fades as a function of $\Delta\tau_i$. Note that the choice to model $a_{i,2}$ statistically, unlike, e.g., the deterministic geometric models in~\cite{wymeersch_beyond_2020}, is made to allow for statistical analysis of the communication reliability.

Given a modulated symbol $\mbf{s} \in \mC^N$, the received baseband signal in the frequency domain from \gls{bs} $i$ across the different subcarriers, $\tilde{\mbf{y}}_i \in \mC^N$, is given by\footnote{We assume identical uplink and downlink channels.
} \cite{wymeersch_beyond_2020}
\begin{align} \label{eq:sysmod}
\tilde{y}_{i,j} =\sqrt{P_{\text{tx}}}\tilde{h}_{i,j}s_j + \tilde{n}_{i,j}
\end{align}
for $j = 0,\dots, N-1$, where $P_{\text{tx}}$ is the transmit power per sub-carrier, $\tilde{n}_{i,j}\sim \mathcal{CN}(0, \sigma_n^2)$ is the noise term with variance $\sigma_n^2$ and $\tilde{h}_{i,j} = a_{i,1}d_j(\tau_{i,1}) + a_{i,2}d_j(\tau_{i,2})$ is the Fourier transform of the channel with $d_j(\tau) = \exp\left(-\jmath\pi2j\Delta_f \tau\right).$

\rev{For later analysis, we note that the channel experiences Rician fading such that the envelope distribution for each subcarrier is Rician with parameters \cite{eggers}
\begin{align}
A_{i,j} & = E[|\tilde{h}_{i,j}|^2] = |E[\tilde{h}_{i,j}]|^2 + \Var[\tilde{h}_{i,j}] \nonumber \\
&= P_L(d_i)\left(1 + \frac{1}{\rho}\exp\left(- \frac{\Delta\tau_i}{\rho} \right) \right),  \label{eq:Rician_mean} \\
K_{i,j} &= \frac{|E[\tilde{h}_{i,j}]|^2}{\Var[\tilde{h}_{i,j}]} = \rho \exp\left(\frac{\Delta\tau_i}{\rho} \right), \label{eq:Rician_K}
\end{align}
where $A_{i,j}$ is the average power and $K_{i,j}$ is the Rician k-factor. The parameters in \eqref{eq:Rician_mean}-\eqref{eq:Rician_K} reveal that the fading distribution does not depend on the subcarrier $j$ although the coefficients $\tilde{h}_{i,0}, \dots, \tilde{h}_{i,N-1}$ are completely dependent since they are driven by the same random variable $a_{i,2}$.}

The system settings used in the examples throughout this letter are summarized in Table \ref{tab:settings}. Note that some of the values are chosen to produce results that clearly show the effect of location uncertainty on reliability with less emphasis on modeling a realistic scenario. 
\begin{table}
    \centering
    \caption{System settings.} \label{tab:settings}
    \begin{tabular}{ccc}
    \textbf{Symbol} & \textbf{Description}  & \textbf{Value} \\
    \hline & & \\[-1.5ex]
    $\begin{bmatrix} x_1 , x_2 \end{bmatrix}$ & BS locations & $\begin{bmatrix} 0 , 1000 \end{bmatrix}$ m \\
    $P_{\text{tx}}$ & Transmit power per sub-carrier & $10$ dBm \\
    $\sigma_n^2$ & Noise variance & $-70$ dBm \\
    $W$ & Bandwidth & $10$ MHz \\
    $f_c$ & Center frequency & $2.1$ GHz \\
    $N$ & Number of sub-carriers & $600$ \\ 
    $\Delta\tau_i$ & Excess delay (same for $i = 1,2$) & $50$ ns \\ 
    $\rho$ & Parameter for power delay profile  & 2 \\
    \end{tabular} \vspace{2mm}
\end{table}
\vspace{-0.75\baselineskip}
\subsection{Localization and communication protocol}
\label{sec:SysModel_B}

The following simple two-step protocol is assumed:

\textbf{1)} When the \gls{ue} turns on for the first time, it estimates its location using a ping --- a single \gls{ofdm} symbol as in \eqref{eq:sysmod} with $s_j = 1$ $\forall$ $j$ --- from each \gls{bs}. To that end, \gls{toa} estimation is employed. Moreover, the pings are used to select the \gls{bs} with which the \gls{ue} will communicate based on the received power, i.e., \gls{bs} $i \in \{1,2\}$ is chosen such that $\|\mbf{\widetilde{y}}_i\|^2$ is maximized. 

\textbf{2)} Once the \gls{ue} has estimated its location and the target \gls{bs}, it starts the communication with the chosen \gls{bs} by sending data (power normalized $E[|s_j|^2]=1$) using the OFDM channel with rate $R$. It is assumed that the channels used in steps 1 and 2 are independent.

To inform rate selection, we introduce the \gls{mar} $R_{\max}$ as an information-theoretic bound, and the \gls{ue} should select $R$ such that it only exceeds $R_{\max}$ with low probability (explained further in Sec. \ref{subsec:capacity}). Following the Sec. \ref{sec:intro}, it is assumed that a mapping between location $x$ and the statistics of $R_{\max}$ is available to the \gls{ue}. We then analyze how localization errors affect the reliability and throughput of the system when the \gls{ue} selects $R$ using different location-based rate selection schemes.


 \label{sec:system}
\section{Statistics of Localization and Communication} \label{sec:theory}
\subsection{Localization} \label{subsec:localization}
\begin{figure}
    \centering
    \includegraphics[width = 0.85\linewidth]{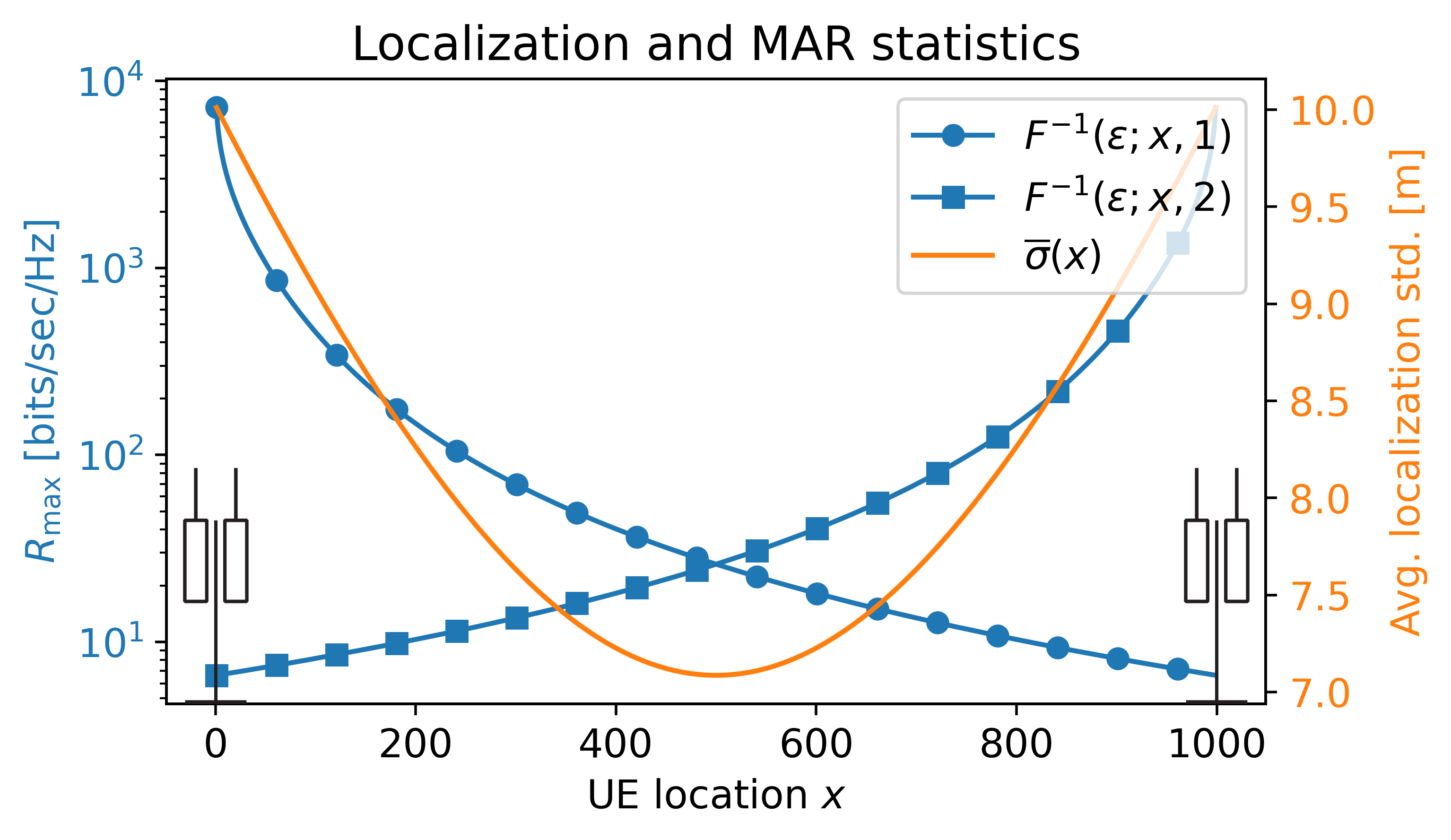}
    \caption{Statistics for localization and MAR. The inverse CDF $F^{-1}(\epsilon; x,i)$ is the $\epsilon$-quantile for $R_{\max}$ at location $x$ when communicating with BS $i$. Here, $\epsilon = 10^{-3}$. $\overline{\sigma}^2(x)$ is the localization variance $\sigma^2(x;\mbs{\phi})$ averaged over $\mbs{\phi}$; the figure shows the standard variation $\overline{\sigma}(x)$.}
    \label{fig:stats}
\end{figure}

In \gls{toa} localization, location is estimated based on the propagation delay of the \gls{los} path, although the accuracy of this method suffers when the \gls{ue} and \gls{bs} clocks are not perfectly synchronized \cite{survey_TOA}.  
We introduce the effect of clock bias $B$ in the localization uncertainty, i.e., the measured delay is $\tilde{\tau}_{i,1} = \norm{x-x_i}/c + B$. Then, given the received signals $\tilde{\mbf{y}}_1, \tilde{\mbf{y}}_2$ from \eqref{eq:sysmod}, 
we use the Cram\'{e}r-Rao inequality to characterize the variance of any unbiased estimator of $x$ as
\begin{align}
    \Var[\hat{x}(\tilde{\mbf{y}}_1, \tilde{\mbf{y}}_2)] \geq J^{-1}(x),
\end{align}
where $J^{-1}(x)$ is the Fisher information corresponding to the location $x$ \cite{Kay}. To find $J(x)$, we first derive the Fisher information with respect to the unknown parameters
\begin{equation} 
\resizebox{.91\linewidth}{!} {$
\mbs{\eta}_i = \begin{bmatrix} \tilde{\tau}_{i,1} & \tilde{\tau}_{i,2} & \mathfrak{R}(a_{i,1}) & \mathfrak{I}(a_{i,1}) & \mathfrak{R}(a_{i,2}) & \mathfrak{I}(a_{i,2}) \end{bmatrix}^{\T} $}
\end{equation}
for $i = 1,2$. For fixed channel coefficients, the normalized received signal $\tilde{\mbf{y}}_i/\sqrt{P_{\text{tx}}}$ follows a circular symmetric, complex Gaussian distribution with mean $\mbs{\mu}(\mbs{\eta}_i) = a_{i,1}\mbf{d}(\tilde{\tau}_{i,1}) + a_{i,2}\mbf{d}(\tilde{\tau}_{i,2})$ and covariance $\frac{\sigma_n^2}{P_{\text{tx}}} \mbf{I}_{N\times N}$. Therefore \cite{Kay}:
\begin{align} \label{eq:fisher_eta}
J(\mbs{\eta}_i) &= \frac{2P_{\text{tx}}}{\sigma_n^2} \sum_{j=0}^{N-1} \mathfrak{R}\left( \frac{\partial \mu_j }{\partial \mbs{\eta}_i}\left(\frac{\partial \mu_j }{\partial \mbs{\eta}_i}\right)^{\H}\right),
\end{align}
whose closed form expression is omitted here due to space limitation. In $\mbs{\eta}_i$, the LoS delay $\tilde{\tau}_{i,1}$ contains information about the location $x$, so we continue with the equivalent Fisher information~\cite{equivalent_fisher} 
\begin{align}
    J^E(\tilde{\tau}_i) = J(\mbs{\eta}_i)_{1,1} - J(\mbs{\eta}_i)_{1,2:6}J^{-1}(\mbs{\eta}_i)_{2:6,2:6}J(\mbs{\eta}_i)_{2:6,1}
\end{align}
where the second term is interpreted as the information loss from the unknown variables. Due to independence of the TOA signals,  $J^E(\tilde{\tau}_{1,1},\tilde{\tau}_{2,1})$ is the diagonal matrix with entries $
J^E(\tilde{\tau}_{1,1}), J(\tilde{\tau}_{2,1})$  and the Fisher information with respect to $(x,B)$ is obtained using the transformation \cite{Kay}
\begin{align} 
    J(x,B) = \mbf{T}^{\T} J^E(\tilde{\tau}_{1,1},\tilde{\tau}_{2,1}) \mbf{T}, \quad \mbf{T} = \begin{bmatrix} 
    \frac{\partial \tilde{\tau}_{1,1}}{\partial x} & \frac{\partial \tilde{\tau}_{1,1}}{\partial B} \\
    \frac{\partial \tilde{\tau}_{2,1}}{\partial x} & \frac{\partial \tilde{\tau}_{2,1}}{\partial B} 
    \end{bmatrix}.
\end{align}
Finally, $J^{-1}(x) = \left(J(x,B)^{-1}\right)_{1,1}$ gives the Cram\'{e}r-Rao lower bound, which is assumed for the variance of $\hat{x}$. 
Additionally, we assume the asymptotic result in which the location estimator $\hat{x}$ follows a Gaussian distribution \cite{madsen_introduction_2011}, that is $\hat{x} \sim \mathcal{N}(x,J^{-1}(x))$. The variance $J^{-1}(x)$ contains the random channel coefficients from the scatter paths, $a_{1,2}$, $a_{2,2}$, and it turns out that only the phases of these, $\phi_{1,2}$, $\phi_{2,2}$, affect the variance while the magnitudes cancel. Denoting $\sigma^2(x;\mbs{\phi}) = J^{-1}(x)$ and using that $\mbs{\phi} = \begin{bmatrix} \phi_{1,2} & \phi_{2,2} \end{bmatrix}^{\T}$ is uniform on $[0,2\pi)^2$, we get the hierarchical model for the output of the localization algorithm
\begin{align} \label{eq:loc_uncer}
    \hat{x} \cond \mbs{\phi} \sim \mathcal{N}(x,\sigma^2(x;\mbs{\phi})), \quad \mbs{\phi} \sim \text{uniform}([0,2\pi)^2). 
\end{align}
For the sake of illustration, Fig.~\ref{fig:stats} shows localization uncertainty for different locations $x$. 
\vspace{-0.75\baselineskip}
\subsection{Rate and Communication Reliability} 
\label{subsec:capacity}
At the physical layer, we assess the system's reliability by its ability to choose a rate $R$ that does not exceed the \gls{mar}. From \eqref{eq:sysmod},  the instantaneous \gls{mar} for the channel between the \gls{ue} at location $x$ and \gls{bs} $i$ is given by \cite{tse_vis}
\begin{align} \label{eq:mar}
    R_{\max}(x,i) = \sum_{j = 0}^{N-1} \log_2\left(1 + \frac{\rev{P_\text{tx}}|\tilde{h}_{i,j}|^2}{\sigma_n^2} \right). 
\end{align}
Fig. \ref{fig:stats} shows statistics for $R_{\max}$ as a function of location $x$. Reliability is  characterized by the outage probability
\begin{equation}
 P(R > R_{\max}(x, i)) = F(R; x,i),
\end{equation}
where $F$ is the \gls{cdf} for $R_{\max}$. Introducing $\epsilon > 0$ as an upper bound for the outage probability, the rate is ideally selected as $R = F^{-1}(\epsilon; x,i)$, also known as the $\epsilon$-outage capacity.
However, if $F$ is not perfectly known, the outage probability is not guaranteed to meet this constraint, and the concept of  \textit{probably correct reliability} measured by the \textit{meta-probability} arises as an approach to characterize the uncertainty\cite{marko}. 

Here, as stated in Sec. \ref{sec:SysModel_B}, it is assumed that the \gls{cdf} has been mapped for all locations prior to transmission, i.e., given a location $x$, $F^{-1}(\epsilon; x,i)$ is perfectly known. Therefore, after estimating its location $\hat{x}$, the \gls{ue} selects the rate using some function $R_{\epsilon,i}(\hat{x})$ (specific examples are introduced in Sec. \ref{sec:rate_select}). 

Given an estimated location $\hat{x}$, the outage probability is
\begin{align}
    p_{\text{out}}(x,\hat{x};i) &= P(R_{\epsilon,i}(\hat{x}) > R_{\max}(x,i) \cond \hat{x}, i) \nonumber  \\
    &= F(R_{\epsilon,i}(\hat{x}); x, i), \label{eq:metaprobability_BS}
\end{align}
and the meta-probability for the link between the \gls{ue} and \gls{bs} $i$ is \cite[Eq. (15)]{marko}
\begin{align} 
    \tilde{p}_{\epsilon}(x; i) &= P\left(P(R_{\epsilon,i}(\hat{x}) > R_{\max}(x,i) \cond \hat{x}, i) > \epsilon\right) \nonumber \\
    &= P_{\hat{x}}(p_{\text{out}}(x,\hat{x};i) > \epsilon). \label{eq:metaprobability}
\end{align}

\begin{figure}[t]
    \centering
    \includegraphics[width = 0.85\linewidth]{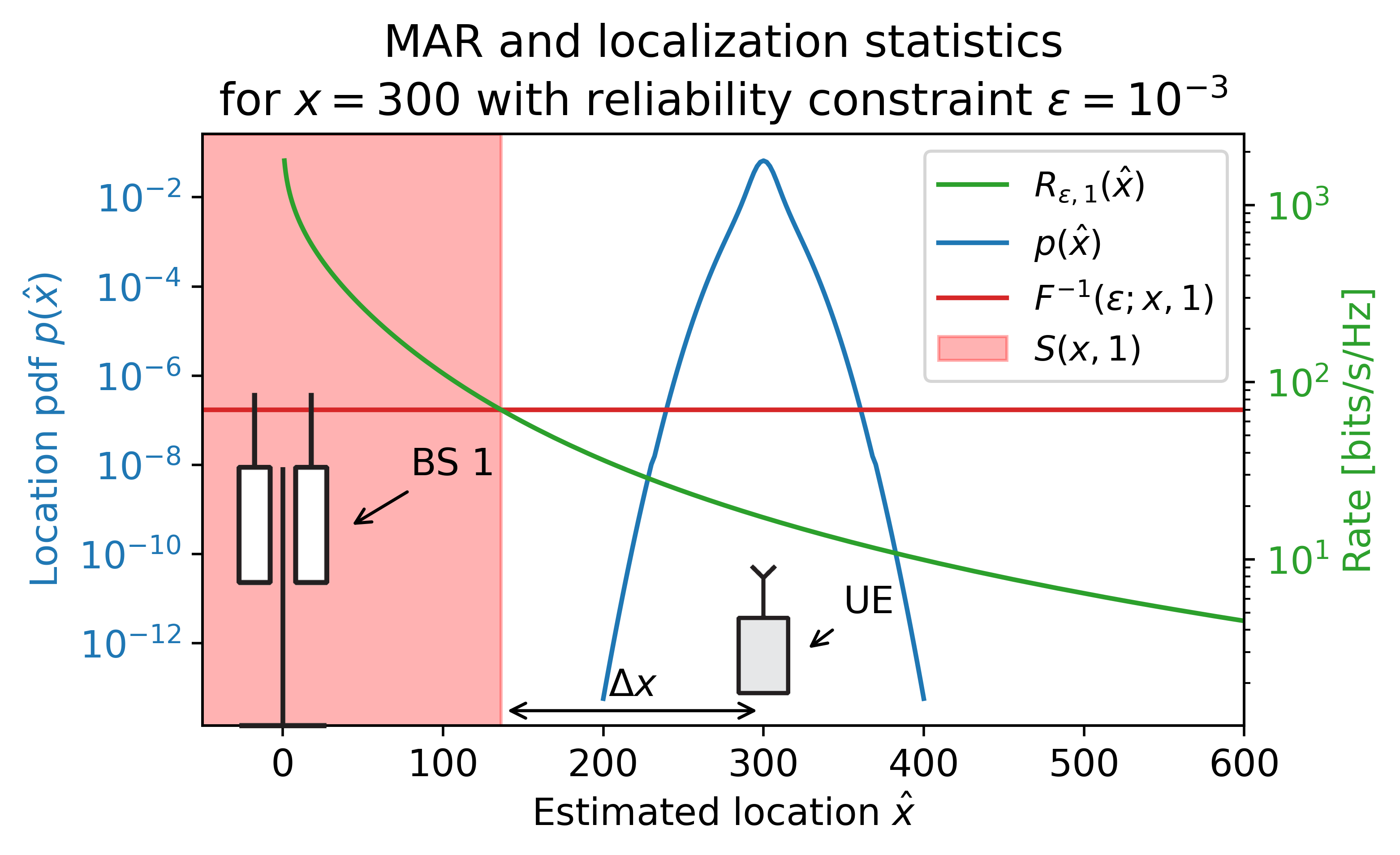}
      \caption{Meta-probability statistics with backoff rate selection function $R_{\epsilon,i}(\hat{x}) = 0.25\cdot F^{-1}(\epsilon; \hat{x},i)$ (see Sec. \ref{subsec:backoff}). The location probability density function (PDF) $p(\hat{x})$ is the marginal PDF of $p(\hat{x},\mbs{\phi})$ according to \eqref{eq:loc_uncer}. The UE location is $300$ m \rev{with $\epsilon$-outage capacity $F^{-1}(\epsilon; x,1) = 70$ bits/s/Hz and the resulting outage region is $S(x,1) = [-136.2,136.2]$ m.}}
    \label{fig:meta_stats}
\end{figure}

Assuming block fading where $R_{\max}(x,i)$ is drawn independently for each block, \eqref{eq:metaprobability} gives the probability that the outage probability exceeds $\epsilon$ for any of these blocks. Averaging over the \glspl{bs} we have
\begin{align} 
    \tilde{p}_{\epsilon}(x) = \sum_{i=1}^2 P_{\hat{x}}(p_{\text{out}}(x,\hat{x};i) > \epsilon)p_i(i; x), \label{eq:p_epsilon}
\end{align}
where $p_i(i; x)$ is the probability of selecting \gls{bs} $i$ which is obtained through Monte-Carlo simulation according to Sec. \ref{sec:SysModel_B}\footnote{The estimated location $\hat{x}$ and selected BS $i$ are dependent since they use the same signal, but their dependence is neglected for the reliability analysis.}. In \eqref{eq:p_epsilon}, the first factor is rewritten by introducing the \textit{outage region}
\begin{align} \label{eq:outage_region}
    S(x,i) = \left\{ \hat{x} \in \mR \cond p_{\text{out}}(x,\hat{x};i) > \epsilon \right\}
\end{align}
such that 
\begin{align} \label{eq:outage_region_probability}
   P_{\hat{x}}(p_{\text{out}}(x,\hat{x};i) > \epsilon) =  P_{\hat{x}}(\hat{x} \in S(x,i)). 
\end{align}
The outage region is interpreted as the region of estimated locations $\hat{x}$ where the rate selection function chooses a rate that is too optimistic for the \gls{mar} at location $x$. 
Fig. \ref{fig:meta_stats} depicts various statistics relevant for the meta-probability using one of the rate selection functions from Sec. \ref{sec:rate_select}. It is observed that the \gls{ue} will choose an overly optimistic rate if it thinks it is closer to the BS than it actually is. The particular rate selection function shown in Fig. \ref{fig:meta_stats} is somewhat conservative; therefore, the outage region is pushed away from the \gls{ue}, and the meta-probability is the probability mass from localization inside the outage region.

Together with the meta-probability, the other metric used to evaluate location-based rate selection methods is the throughput ratio, defined as the ratio \cite{marko} 
\begin{align}
   \omega_{\epsilon}(x) &= \frac{E\left[R_{\epsilon,i}(\hat{x})\mathbbm{1}\{ R_{\epsilon,i}(\hat{x}) \leq R_{\max}(x,i)\} \right]}{E[ R^*_{\epsilon,i}(x)\mathbbm{1}\{ R^*_{\epsilon,i}(x)\leq R_{\max}(x,i)\}]},
\end{align}
between the throughput using $R_{\epsilon,i}(\hat{x})$ and the optimal throughput using $R^*_{\epsilon,i}(x) = F^{-1}(\epsilon;x,i)$ where $\mathbbm{1}$ is the indicator function. 
The throughput ratio is expanded with repeated use of the law of total expectation, yielding
\begin{equation} \label{eq:throughput_expanded} 
\omega_{\epsilon}(x) =  \dfrac{E_i\left[E_{\mbs{\phi}} \left[ E_{\hat{x}\cond \mbs{\phi}} \left[R_{\epsilon,i}(\hat{x})\left(1 - p_{\text{out}}(x,\hat{x};i) \right) \cond i, \mbs{\phi} \right] \cond i \right]\right]}{E_i[R^*_{\epsilon,i}(x)](1 - \epsilon)}.
\end{equation}

\section{Location-Aware Rate Selection} \label{sec:rate_select}
Location-based rate selection methods naturally depends on the rate function  $R_{\epsilon,i}(\hat{x})$, and the ultimate goal would be to solve the optimization problem
\begin{align} \label{eq:optimize}
    \sup_{R_{\epsilon,i}} \int \omega_{\epsilon}(x) \ dx, \quad \text{s.t.} \quad  \tilde{p}_{\epsilon}(x) \leq \delta \ \forall x,
\end{align}
where $\delta$ is the \textit{confidence parameter} bounding the meta-probability. This section introduces three examples of rate selection functions that account for uncertainty in the location estimate by selecting a conservative rate compared to the optimal rate selection function, i.e., $R_{\epsilon,i}(\hat{x}) < R_{\epsilon,i}^*(\hat{x})$. We limit the search to functions that satisfy the inequality constraint in \eqref{eq:optimize} and then analyze the tradeoffs between reliability and throughput. 

\subsection{Backoff rate selection} \label{subsec:backoff}
The backoff rate selection function chooses the rate proportional to the $\epsilon$-quantile for the \gls{mar} at the estimated location: 
\begin{align}
R_{\epsilon,i}(\hat{x}) = k \cdot F^{-1}(\epsilon ; \hat{x},i),    
\end{align}
where $0 < k \leq 1$ is the proportionality constant. The parameter $k$ is interpreted as how conservatively the system selects the rate relative to the optimal selection when the location is perfectly known. Finding $k$ such that the meta-probability is below the confidence parameter $\delta$ requires knowledge of the  system statistics, including location uncertainty, which may or may not be available in practice. For the sake of illustration, we simply choose the maximum $k \in (0,1]$ such that the meta-probability is below $\delta$ within a range of locations. 

\subsection{Confidence intervals rate selection}
This approach considers a confidence interval for estimated location and then chooses the minimum rate within that interval. Denoting $\text{CI}_{\alpha}(\hat{x})$ as the confidence interval for $x$ with confidence level $(1-\alpha)$, the rate is selected as
\begin{align}
    R_{\epsilon,i}(\hat{x}) = \min_{x} \left\{ F^{-1}(\epsilon; x,i)  \cond x \in \text{CI}_{\alpha}(\hat{x}) \right\}.
\end{align} 
Setting an appropriate confidence level and obtaining the confidence interval again requires knowledge of the system statistics. Similarly to the backoff method, we find the appropriate $\alpha \in (0,1)$ considering the constraint on the meta-probability. Specifically, we use the approximate interval\footnote{The confidence interval in \eqref{eq:interval} is only approximate since it assumes that $\hat{x}$ is Gaussian, where in reality, it is only conditionally Gaussian.}
\begin{equation}
    \text{CI}_{\alpha}(\hat{x}) = \left[\hat{x} - q_{1-\alpha/2} \sigma(\hat{x}), \hat{x} + q_{1-\alpha/2} \sigma(\hat{x})\right], \label{eq:interval} 
\end{equation} 
where $q$ are the quantiles of the standard Gaussian distribution and $\sigma(\hat{x})$ is the standard deviation for localization at $\hat{x}$ according to \eqref{eq:loc_uncer}, which is assumed to be known. 
\subsection{Oracle rate selection}
Lastly, \textit{oracle rate selection} is introduced, which attempts to solve \eqref{eq:optimize} by exhaustive search for the selected rate. This obviously requires a full statistical characterization of the system and is not of much practical interest. However, it serves as an upper bound and importantly shows that throughput suffers due to localization uncertainty even when the statistics of the system are fully known. 

\rev{\section{Analytical results for reliability}}
\rev{The proposed meta-probability in \eqref{eq:metaprobability_BS}-\eqref{eq:outage_region_probability} is difficult to analyze and does not have a closed-form expression. This section will provide a few analytical results to lessen the numerical evaluation and also provide some intuition for the meta-probability, i.e., reliability.}

\rev{To ease the numerical evaluation of \eqref{eq:outage_region_probability}, we  analyze the outage region in \eqref{eq:outage_region} for the aforementioned rate selection functions}. Interestingly, we observe that the region reduces to a single interval for the considered 1-D scenario, i.e., $S(x,i) = [x_{\min,i},x_{\max,i}]$. Thus, according to the hierarchical model from \eqref{eq:loc_uncer}, we can expand \eqref{eq:outage_region_probability} as
\begin{align}
     &P_{\hat{x}}(\hat{x} \in S(x,i)) = \frac{1}{4\pi^2}\int_{[0,2\pi)^2}  P_{\hat{x}\cond \mbs{\phi}}(\hat{x} \in S(x,i) \cond \mbs{\phi}) \ d\mbs{\phi} \notag  \\ 
     &=  \frac{1}{4\pi^2}\int_{[0,2\pi)^2} Q\left(\frac{x_{\min,i} - x}{\sigma(x;\mbs{\phi})}\right) - Q\left(\frac{x_{\max,i} - x}{\sigma(x;\mbs{\phi})}\right) \ d\mbs{\phi}, \label{eq:meta3}
\end{align}
using the $Q$-function for the standard Gaussian distribution.

\rev{To get some intuition for how location affects reliability, consider the simplified case with only a single subcarrier ($N = 1$) and rate selection using the backoff approach for some choice of $k$. We shall see that this setting allows us to derive a closed-form expression for the $\epsilon$-outage capacity by relying on a few approximations. First, assume without loss of generality that the \gls{ue} at location $x > 0$ is communicating with \gls{bs} $1$ ($i = 1$) at location $x_{\text{BS}} = 0$ such that the distance between transmitter and receiver is simply $d = x$. Focusing on the ultra-reliable regime,  
the lower tail of the received power can be modeled  using a power-law approximation. Dropping the subscripts, the received power is denoted $Y = |\tilde{h}|^2$ , and the power tail approximation for the Rician fading channel is \cite{eggers}
\begin{align}
F_Y(y) &= P(|\tilde{h}|^2 \leq y) \nonumber \\
&\approx \frac{y}{A}(1+K)e^{-K} \quad (\text{when } y \approx 0) \nonumber \\
&=  y x^2 \underbrace{\frac{16\pi^2}{\lambda^2}\rho e^{\Delta\tau/\rho + \rho e^{\Delta\tau/\rho}}}_{\psi} \label{eq:cdf_deriv},
\end{align}
where \eqref{eq:cdf_deriv} uses the expressions for $A$ and $K$ given in \eqref{eq:Rician_mean}-\eqref{eq:Rician_K}. For one subcarrier, the \gls{mar} reduces to a single term in \eqref{eq:mar}, and its \gls{cdf} follows directly from \eqref{eq:cdf_deriv} as
\begin{align}
    F(R; x) = P(R \geq R_{\max}; x) = \left(2^R - 1\right)\frac{\sigma_n^2}{P_{\text{tx}}} x^ {-2}\psi,
\end{align}
and the $\epsilon$-outage capacity becomes
\begin{align}
    R_{\epsilon}(x) = F^{-1}(\epsilon; x) = \log_2\bigg(1 + x^{-2}\underbrace{\frac{P_{\text{tx}}\epsilon}{\sigma_n^2 \psi}}_{\psi'}\bigg) \label{eq:outage_analytical}
\end{align}
With backoff rate selection, the target outage probability is then exceeded whenever $k \cdot F^{-1}(\epsilon; \hat{x}) > F^{-1}(\epsilon; x)$ depending on the estimated location $\hat{x}$ as characterized by the outage region in \eqref{eq:outage_region}. Denoting the localization error as $\Delta x = x - \hat{x}$, the edges of the outage region are given for those values of $\Delta x$ such that
\begin{align}
    k \cdot F^{-1}(\epsilon; x - \Delta x) = F^{-1}(\epsilon; x), \label{eq:outage_condition}
\end{align}
and the closest edge to the \gls{ue} is hence in the region $0 < \hat{x} < x$ 
(see Fig. \ref{fig:meta_stats}). Combining \eqref{eq:outage_analytical} and \eqref{eq:outage_condition}, it follows that
\begin{align}
\Delta x = x - \sqrt{\frac{\psi'}{\left(1 - x^{-2}\psi' \right)^{1/k}-1}} \label{eq:deltaXcomplicated}
\end{align}
for the edge closest to the \gls{ue}. Eq. \eqref{eq:deltaXcomplicated} does not provide any immediate insight, however, using the approximation ${\log_2(1 + y) \approx y/\ln(2)}$ for $y \approx 0$, we get
\begin{align}
   \Delta x \approx x(1-\sqrt{k}). \label{eq:deltaesay}
\end{align}
Observing Fig. \ref{fig:meta_stats}, we see that the outage probability increases as $\Delta x$ decreases, and \eqref{eq:deltaesay} now reveals that $\Delta x$ decreases in $k$. When $k \to 1$, the rate is selected less conservatively, and the distance goes to $0$. Interestingly, it is also seen that $\Delta x$ scales linearly with $x$. As such, a larger localization error is allowed as the \gls{ue} moves farther away from the \gls{bs}, which may result in increased reliability depending on how the localization error changes. Finally, it is observed that \eqref{eq:deltaesay} does not depend on any of the factors in $\psi'$, which means that the outage probability is essentially invariant to the system parameters as long as $\epsilon$ is sufficiently low for the approximations to hold\footnote{The error between the expressions for $\Delta x$ in \eqref{eq:deltaXcomplicated} and \eqref{eq:deltaesay} is negligible for the system settings in Table \ref{tab:settings}. In general, the approximation in \eqref{eq:deltaesay} is accurate except when $x \approx 0$, $k \approx 0$ and/or $\psi' \to \infty$.}. The following section will show how these intuitions hold in the general case with multiple subcarriers.}  
\section{Evaluation of reliability and throughput} \label{sec:results}
\begin{figure}
    \centering
    \includegraphics[width = 0.85\linewidth]{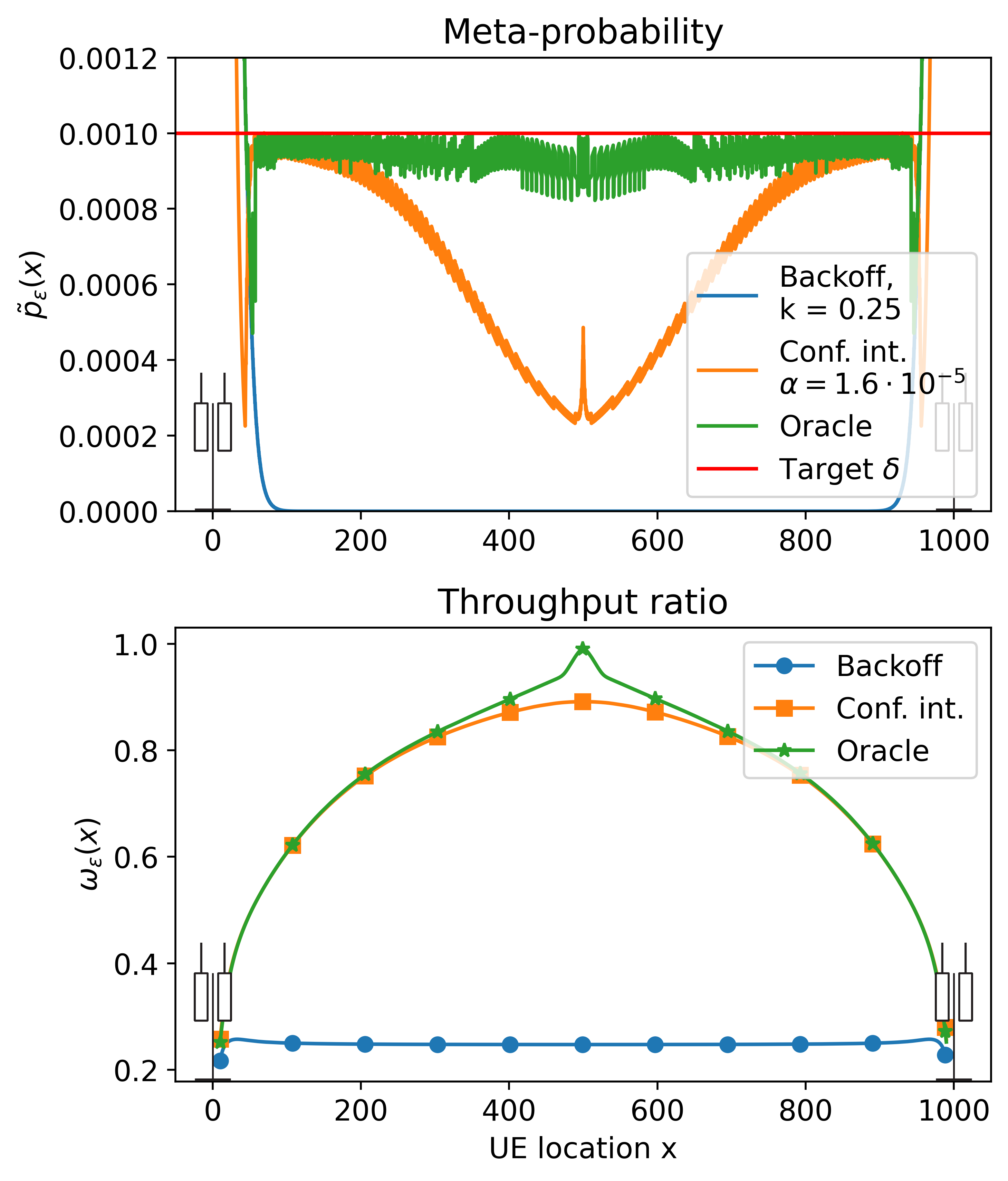}
    \caption{Meta-probability and throughput ratio of the different rate selection schemes for each UE location $x \in [10,990]$. Backoff rate selection uses $k = 0.25$ and confidence interval rate selection uses $\alpha = 1.6\cdot 10^{-5}$.}
    \label{fig:meta}
\end{figure}

The meta-probability in \eqref{eq:p_epsilon} and throughput ratio in \eqref{eq:throughput_expanded} are now evaluated under the different rate selection schemes from Sec. \ref{sec:rate_select} and settings in Table \ref{tab:settings}. The rate selection functions are calibrated such that the meta-probability with  $\epsilon = 10^{-3}$ is below $\delta = 10^{-3}$ for $x \in [45,955]$. Fig. \ref{fig:meta} depicts the results for the meta-probability and throughput ratio. Both \eqref{eq:throughput_expanded} and \eqref{eq:meta3} are evaluated through numerical integration.

The results for the backoff and confidence interval methods in Fig. \ref{fig:meta} show that the meta-probability, i.e., the probability of selecting a rate that exceeds the \gls{mar}, tends to decrease when the \gls{ue} is farther away from the \glspl{bs}, which \rev{aligns well with the previous intuition}. Two effects contribute to this: the decreasing location uncertainty and the rapidly decreasing $\epsilon$-quantile for $R_{\max}$ as the distance between the \gls{ue} and the \gls{bs} increases  (see Fig. \ref{fig:stats}). To understand the latter, note that the two methods select rates based on the $\epsilon$-quantile. Therefore, even a small localization error close to a \gls{bs} can cause the \gls{ue} to pick a much higher rate, whereas the same error farther away causes a smaller change in the selected rate. 
In fact, considering that the changes in the average location uncertainty are almost negligible compared to the rapid variations in the $\epsilon$-quantile (see Fig. \ref{fig:stats}), it is mainly the change in the quantile that causes the meta-probability to decrease, hence reliability to increase, with the distance to the \gls{bs}\footnote{{This has been numerically verified by computing the meta-probability under constant $\Var[\hat{x}]$, where we see similar curves as in Fig. \ref{fig:meta}.}}. For oracle rate selection, we see values close to the confidence parameter $\delta$, as expected. \rev{Regarding the throughput ratio in Fig. \ref{fig:meta}, it is interesting to see that for the backoff approach it }is more or less flat with $\omega_{\epsilon}(x) \approx k$.
In contrast, the throughput ratios for the confidence interval and oracle approach have a strong dependence on $x$. For these methods, we also observe that the throughput ratios increase farther away from the \glspl{bs}, again explained by the $\epsilon$-quantile for $R_{\max}$. To see this, consider the extreme case where the distribution $F(R;x,i)$ is constant for all locations $x$. Here, the \gls{ue} would choose $R = F^{-1}(\epsilon;\hat{x},i)$, which is invariant to the estimated location; thus, the meta-probability is zero, and the optimal throughput ratio is achieved. In our setup, when the \gls{ue} is far away from the \glspl{bs}, it experiences a similar case where the distribution is almost constant within the range of likely estimated locations (see Fig. \ref{fig:stats}), thus enabling the \gls{ue} to be less conservative and achieve higher throughput. 

In summary, we observe degradation in the system when the \gls{ue} is close to a \gls{bs} and vice versa due to how quickly the $\epsilon$-quantile for $R_{\max}$ changes for different $x$, leading to the conclusion that lower but more spatially consistent channel statistics are desirable for location-based rate selection. 
In fact, we arrive at the fundamental observation that the spatial variation of channel statistics determines the quality of location as a proxy for reliability, which generalizes to other channels, number of antennas, higher dimension for localization, etc.


\section{Conclusions} \label{sec:conclusion}

This letter has analyzed the impact of location uncertainty on communication reliability through a rigorous statistical framework. By eliminating all other sources of uncertainty, we have shown that localization error alone considerably impacts the reliability, especially in areas where channel statistics rapidly varies in space, e.g., close to the \glspl{bs}. Conservative rate selection schemes can avoid this at the expense of reduced throughput. Different rate selection functions were considered, but ultimately this task requires accurate knowledge of channel statistics at different locations to ensure a certain level of reliability. Thus, we have shown that it is not straightforward to use uncertain location as a reliability proxy. This letter analyzed a simple setting, and the followup work will consider extensions such as higher dimensions for localization.

\ifCLASSOPTIONcaptionsoff
  \newpage
\fi

\bibliographystyle{ieeetr}
\bibliography{references}


\end{document}